\documentstyle[prc,aps,psfig,amsmath]{revtex}
\begin{document}

\draft

\twocolumn [ \hsize\textwidth\columnwidth\hsize\csname
@twocolumnfalse\endcsname

\title{Three-potential formalism for the three-body 
scattering problem with attractive Coulomb interactions}

\author{Z.\ Papp${}^{1,2}$, C-.Y.\ Hu${}^{1}$,  Z.\ T.\ Hlousek${}^{1}$, 
B.~K\'onya${}^{2}$ and S.\ L.\ Yakovlev${}^{3}$ }
\address{${}^{1}$ Department of Physics and Astronomy, 
California State University, Long Beach, CA 90840, USA \\
${}^{2}$ Institute of Nuclear Research of the
Hungarian Academy of Sciences, Debrecen, Hungary \\
${}^{3}$ Department of Mathematical and Computational Physics, 
St.\ Petersburg State University, St.\ Petersburg,  Russia }
\date{\today}
\maketitle

\begin{abstract}
\noindent
A three-body scattering process in the presence of Coulomb interaction
can be decomposed formally into a two-body single channel,
a two-body multichannel and a genuine three-body scattering.
The corresponding integral equations are coupled Lippmann-Schwinger and 
Faddeev-Merkuriev integral equations. We solve them by applying 
the Coulomb-Sturmian separable expansion method.  We present elastic 
scattering and reaction cross sections of the $e^++H$ system both below and
above the $H(n=2)$ threshold. We found excellent agreements with previous
calculations in most cases.
\end{abstract}

\vspace{0.5cm} 
\pacs{PACS number(s): 31.15.-p, 34.10.+x, 34.85.+x, 21.45.+v, 03.65.Nk, 
02.30.Rz, 02.60.Nm}
]

The three-body Coulomb scattering
problem is one of the most challenging long-standing problems
of non-relativistic quantum mechanics. The source of the difficulties
is related to the long-range character of the Coulomb potential.
In the standard scattering theory it is
supposed that the particles move freely  asymptotically. That is
not the case if Coulombic interactions are involved.
As a result the fundamental equations of the three-body
problems, the Faddeev-equations, become ill-behaved if they are applied
for Coulomb potentials in a straightforward manner.

The first, and formally exact, approach was proposed by Noble \cite{noble}. 
His formulation was designed for solving the nuclear three-body
Coulomb problem, where all Coulomb interactions are repulsive.
The interactions were split into short-range and long-range Coulomb-like
parts and the long-range parts were formally included in the 
"free" Green's operator.
Therefore the corresponding Faddeev-Noble equations become mathematically
well-behaved and in the absence of Coulomb interaction 
they fall back to the standard equations. However, the associated 
Green's operator is not known. This formalism, as presented
at that time, was not suitable for practical calculations.

In Noble's approach the separation of the Coulomb-like potential into
short-range and long-range parts were carried out in the 
two-body configuration space.
Merkuriev extended the idea of Noble by performing the 
splitting in the three-body configuration space.
This was a crucial development since it made possible to treat attractive
Coulomb interactions on an equal footing as repulsive ones. 
This theory has been developed using integral equations with
connected (compact) kernels and transformed 
into  configuration-space differential
equations with asymptotic boundary conditions \cite{fm-book}. 
In practical calculations, so far 
only the latter version of the theory has been considered. The primary
reason is that the more complicated structure of the Green's operators 
in the kernels of the
Faddeev-Merkuriev integral equations has not yet allowed any direct solution.
However, use of integral equations is a very appealing approach since
no boundary conditions are required.

Recently, one of us has developed a novel method 
for treating the three-body problem with repulsive
Coulomb interactions in three-potential picture
\cite{pzsc}. In this approach a three-body Coulomb scattering
process can be decomposed formally into a two-body single channel,
a two-body multichannel and a genuine three-body scattering.
The corresponding integral equations are coupled Lippmann-Schwinger
and  Faddeev-Noble integral equations, which were solved by  using
the Coulomb-Sturmian separable expansion method.
The approach was tested first
for bound-state problems \cite{pzwp} with repulsive Coulomb plus nuclear
potential. Then it was extended to calculate $p-d$
scattering at energies below the breakup threshold \cite{pzsc} and
more recently we have used the method  to calculate 
resonances  of three-$\alpha$ systems \cite{zis}. 
Also atomic  bound-state problems with attractive Coulomb
interactions have been considered \cite{pzatom}.
These calculations showed an excellent agreement with the results of other 
well established methods. The efficiency and the accuracy 
of the method was demonstrated. 

The aim of this paper is to generalize this method for solving the three-body
Coulomb problem with repulsive and attractive Coulomb interactions.
We combine the concept of three-potential formalism with the Merkuriev's
splitting of the interactions and solve the resulting set of 
Lippmann-Schwinger and Faddeev-Merkuriev integral equations by applying the 
Coulomb-Sturmian separable expansion method. In this paper we restrict 
ourselves to energies below the three-body breakup threshold.

\section{Integral equations of the three-potential picture}

We consider a three-body system with  Hamiltonian 
\begin{equation}
H=H^0 + v_\alpha^C + v_\beta^C + v_\gamma^C,
\label{H}
\end{equation}
where $H^0$ is the three-body kinetic energy 
operator and $v_\alpha^C$ denotes the Coulomb-like
interaction in subsystem $\alpha$. The potential
$v_\alpha^C$ may have repulsive
or attractive Coulomb tail and any short-range component.
We use the usual
configuration-space Jacobi coordinates
 $x_\alpha$ and $y_\alpha$; $x_\alpha$ is the coordinate
between the pair $(\beta,\gamma)$ and $y_\alpha$ is the
coordinate between the particle $\alpha$ and the center of mass
of the pair $(\beta,\gamma)$.
Thus the potential $v_\alpha^C$, the interaction of the
pair $(\beta,\gamma)$, appears as $v_\alpha^C (x_\alpha)$.
We also use the notation $X=\{x_\alpha,y_\alpha \}\in{\bf R}^6$.

\subsection{Merkuriev's cut of the Coulomb potential}

The  Hamiltonian (\ref{H}) is defined in the three-body 
Hilbert space. The two-body potential operators are formally
embedded in the three-body Hilbert space
\begin{equation}
v^C = v^C (x) {\bf 1}_{y}.
\label{pot0}
\end{equation}
Merkuriev introduced a separation of the three-body
configuration space into different
asymptotic regions. The two-body asymptotic region $\Omega_\alpha$ is
defined as a part of the three-body configuration space where
the conditions
\begin{equation}
|x_\alpha| <  x^0_\alpha ( 1  +
|y_\alpha|/ y^0_\alpha)^{1/\nu},
\label{oma}
\end{equation}
with $x^0_\alpha, y^0_\alpha >0$ and $\nu > 2$, are satisfied.
He proposed to split the Coulomb interaction  in 
the three-body configuration space into
short-range and long-range terms 
\begin{equation}
v_\alpha^C =v_\alpha^{(s)} +v_\alpha^{(l)} ,
\label{pot}
\end{equation}
where the superscripts
$s$ and $l$ indicate the short- and long-range
attributes, respectively. 
The splitting is carried out with the help of a splitting function $\zeta$,
\begin{eqnarray}
v^{(s)} (x,y) & = & v^C(x) \zeta (x,y),
\\
v^{(l)} (x,y) & = & v^C(x) [1- \zeta (x,y) ].
\label{potl}
\end{eqnarray}
The function $\zeta$ is defined such that 
\begin{equation}
\zeta(x,y) \xrightarrow{X \to \infty}
\left\{ 
\begin{array}{ll}
1, &  X \in \Omega_\alpha \\
0  & \mbox{otherwise.}
\end{array}
\right.
\end{equation}
In practice, in the configuration-space differential equation
approaches, usually the functional form
\begin{equation}
\zeta (x,y) =  2/\left\{1+ \left[ {(x/x^0)^\nu}/{(1+y/y^0)} \right] \right\},
\label{oma1}
\end{equation}
was used.

The long-range Hamiltonian is defined as
\begin{equation}
H^{(l)} = H^0 + v_\alpha^{(l)}+ v_\beta^{(l)}+ v_\gamma^{(l)},
\label{hl}
\end{equation}
and its resolvent operator is
\begin{equation}
G^{(l)}(z)=(z-H^{(l)})^{-1}.
\end{equation}
Then, the three-body Hamiltonian takes the form
\begin{equation}
H = H^{(l)} + v_\alpha^{(s)}+ 
 v_\beta^{(s)}+ v_\gamma^{(s)}.
\label{hll}
\end{equation}

In the conventional Faddeev theory the wave function 
components are defined by
\begin{equation}
| \psi_\alpha \rangle  = (z-H^0)^{-1} v_\alpha | \Psi \rangle,
\label{fagyi}
\end{equation}
where $v_\alpha$ is a short-range potential 
and $| \psi_\alpha \rangle$ is the Faddeev
component of the total wave function $| \Psi \rangle$. 
While the total wave function $| \Psi \rangle$,
in general, has three different kind of two-body asymptotic channels,
$| \psi_\alpha \rangle$ possesses only 
$\alpha$-type two-body asymptotic channel.
The other channels are suppressed by the short-range potential
$v_\alpha$. This procedure is called asymptotic filtering and it guarantees
the asymptotic orthogonality of the Faddeev components  \cite{vanzani}.

\begin{figure}
\psfig{file=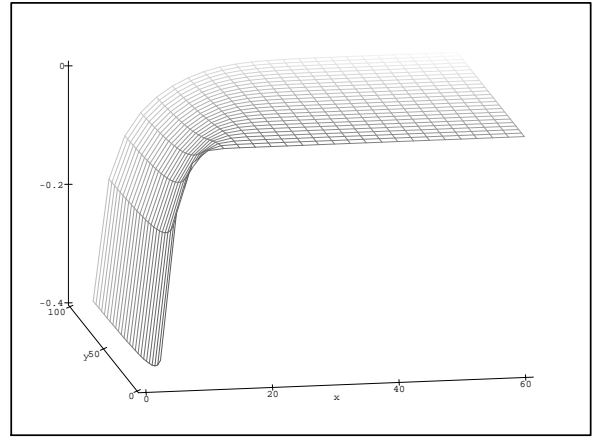,width=8.5cm,angle=-90}
\caption{The short-range part $v^{(s)}$ of the $-1/x$ attractive
Coulomb potential. }
\label{vs}
\end{figure}

\begin{figure}
\psfig{file=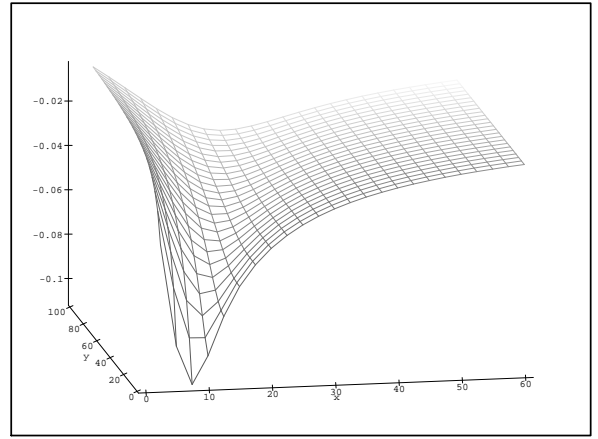,width=8.5cm,angle=-90}
\caption{The long range part $v^{(l)}$ of the $-1/x$ attractive
Coulomb potential. }
\label{vl}
\end{figure}

The aim of the Merkuriev procedure was to formally obtain a three-body 
Hamiltonian with short-range potentials $v^{(s)}$ and long-range
Hamiltonian $H^{(l)}$ in order that we can repeat the procedure of 
the conventional Faddeev theory.
The total wave function 
$|\Psi  \rangle $ is split into three components,
\begin{equation}
|\Psi  \rangle = |\psi_{\alpha} \rangle + |\psi_{\beta} \rangle +
|\psi_{\gamma} \rangle,
\end{equation}
with components defined by 
\begin{equation}
|\psi_{\alpha} \rangle= G^{(l)} v_\alpha^{(s)} |\Psi  \rangle.
\label{fdec}
\end{equation}
This procedure is an example of asymptotic filtering. The short-range potential
 $v_\alpha^{(s)}$ acting on $|\Psi  \rangle$ 
suppresses the possible $\beta$ and $\gamma$ asymptotic two-body
channels, provided $G^{(l)}$ itself does not introduce any new two-body
asymptotic channels.  With the Merkuriev splitting this is avoided because
$H^{(l)}$ does not have two-body asymptotic channels even if some 
of the long-range potentials have attractive Coulomb tail. 
In the attractive case  $v^{(l)}$ appears as
a valley along the $y=x^\nu$ parabola-like
curve with Coulomb-like asymptotic behavior in
$x$ at any finite $y$. (See Figs.\ \ref{vs} and \ref{vl} for the short- 
and long-range parts, respectively). 
However, as $y \to \infty$ the depth of the
valley goes to zero, consequently the two-body bound states are pushed up, and 
finally the system does not have any two-body asymptotic channels.
We note that the Merkuriev formalism contains the Noble's  in the limit 
$y^0\to\infty$.

\subsection{The three-potential picture }

In Ref.\ \cite{pzsc} the three body scattering problem with repulsive 
Coulomb interactions were considered
in the three-potential picture. In this picture the 
scattering process  can be decomposed formally
into three consecutive scattering processes:
a two-body single channel, a two-body multichannel and a genuine three-body
scattering. This formalism also provides the integral equations and the 
method of constructing the $S$-matrix.
Below we adapt this formalism to attractive Coulomb interactions along the
Merkuriev approach.

The asymptotic Hamiltonian is defined as 
\begin{equation}
H_\alpha=H^0 +  v_\alpha^C,
\end{equation}
and the asymptotic states are the eigenstates of $H_\alpha$ 
\begin{equation}
H_\alpha | \Phi_{\alpha} \rangle = E | \Phi_{\alpha} \rangle,
\end{equation}
where $\langle x_\alpha y_\alpha |
 \Phi_{\alpha} \rangle = \langle
y_\alpha| \chi_{\alpha} \rangle \langle 
x_\alpha | \phi_{\alpha} \rangle$ 
is a product of a scattering state in coordinate
$y_\alpha$ and a bound state in the two-body subsystem $x_\alpha$.

We define the two asymptotic long-range Hamiltonians as 
\begin{equation}
H_\alpha^{(l)}=H^0 + v_\alpha^{C}  + v_\beta^{(l)} +v_\gamma^{(l)}
\end{equation}
and 
\begin{equation}
\widetilde{H}_\alpha=H^0 + v_\alpha^{C} + u_\alpha^{(l)},
\label{htilde}
\end{equation}
where $u_\alpha^{(l)}$ is an auxiliary potential in 
coordinate $y_\alpha$,
and it is required to have the asymptotic form 
\begin{equation}
u_\alpha^{(l)} \sim  {Z_\alpha (Z_\beta+Z_\gamma) }/{y_\alpha}
\end{equation}
as ${y_\alpha \to \infty}$. In fact, $u_\alpha^{(l)}$ 
is an effective Coulomb-like
interaction between the center of mass of the subsystem 
$\alpha$ (with
charge $Z_\beta+Z_\gamma$) and the third particle 
(with charge $Z_\alpha$). We introduced this potential in order that
we compensate the long range Coulomb
tail of $v_\beta^{(l)} +v_\gamma^{(l)}$ in $\Omega_\alpha$.

Let us introduce the resolvent operators: 
\begin{equation}
G(z)=(z-H)^{-1},
\end{equation}
\begin{equation}
G_\alpha ^{(l)}(z)=(z-H_\alpha^{(l)})^{-1},
\end{equation}
\begin{equation}
\widetilde{G}_\alpha (z)=(z-\widetilde{H}_\alpha )^{-1}.
\end{equation}
The operator $G_\alpha^{(l)}$ is the long-range channel 
Green's operator
and $\widetilde{G}_\alpha $ is the channel 
distorted long-range Green's
operator. 
These operators are connected via
the following resolvent relations: 
\begin{equation}
G(z)=G_\alpha^{(l)}(z)+G_\alpha^{(l)}(z) V^\alpha G(z),  \label{g3b}
\end{equation}
\begin{equation}
G_\alpha^{(l)}(z)=\widetilde{G}_\alpha (z)+
\widetilde{G}_\alpha (z)U^\alpha
G_\alpha^{(l)}(z),  \label{g2b}
\end{equation}
where $V^\alpha =v^{(s)}_\beta +v^{(s)}_\gamma $ and 
$U^\alpha =v_\beta^{(l)}+v_\gamma^{(l)}-u_\alpha ^{(l)}$.

The scattering state, which evolves from the asymptotic
state $|\Phi_\alpha \rangle$ under the influence of
$H$, is given as
\begin{equation}
| \Psi_\alpha^{(\pm)} \rangle = 
\lim\limits_{\varepsilon\to 0} \mbox{i}
\varepsilon G (E_\alpha \pm \mbox{i}\varepsilon)| 
\Phi_\alpha \rangle.
\label{Psidef}
\end{equation}
Similarly, we can define the following auxiliary scattering states
\begin{equation}
| \Phi_\alpha^{{(l)}(\pm)} \rangle = 
\lim\limits_{\varepsilon\to 0} \mbox{i}
\varepsilon G_\alpha^{(l)} (E\pm \mbox{i}\varepsilon) |
 \Phi_\alpha \rangle \label{phil}
\end{equation}
and
\begin{equation}
| \widetilde{\Phi}_{\alpha}^{(\pm)}\rangle = 
\lim\limits_{\varepsilon\to 0} 
\mbox{i}\varepsilon \widetilde{G}_\alpha
 (E\pm \mbox{i}\varepsilon) |
\Phi_{\alpha} \rangle,  \label{phitild}
\end{equation}
which describe scattering processes due to Hamiltonians
$H^{(l)}_\alpha$ and $\widetilde{H}_\alpha$, respectively.

The S-matrix elements of scattering processes are obtained from the
resolvent of the total Hamiltonian by the reduction technique \cite{redu} 
\begin{equation}
S_{\beta j,\alpha i}=
\lim\limits_{t\to \infty }\lim\limits_{\varepsilon \to
0}\mbox{i}\varepsilon \mbox{e}^{\mbox{i}(E_{\beta j}-
E_{\alpha i})t}\langle
\Phi _{\beta j}|G(E_{\alpha i}+\mbox{i}\varepsilon )|
\Phi _{\alpha i}\rangle .  \label{sm}
\end{equation}
The subscript $i$ and $j$ denotes the $i$-th and 
$j$-th eigenstates of the
corresponding subsystems, respectively. 
If we substitute (\ref{g3b}) into (\ref{sm}) we  get
the following two terms:
\begin{equation}
S_{\beta j,\alpha i}^{(1,2)}=\lim\limits_{t\to \infty
}\lim\limits_{\varepsilon \to 0}\mbox{i}\varepsilon 
\mbox{e}^{\mbox{i}
(E_{\beta j}-E_{\alpha i})t}\langle \Phi _{\beta j}|
G_\alpha^{(l)}(E_{\alpha i}+
\mbox{i}\varepsilon )|\Phi _{\alpha i}\rangle   \label{s12}
\end{equation}
\begin{eqnarray}
S_{\beta j,\alpha i}^{(3)}&=&\lim\limits_{t\to \infty
}\lim\limits_{\varepsilon \to 0}\mbox{i}\varepsilon \mbox{e}^{\mbox{i}
(E_{\beta j}-E_{\alpha i})t}\langle \Phi _{\beta j}|
G_\alpha^{(l)}(E_{\alpha i}+\mbox{i}\varepsilon ) \nonumber \\
&& V^\alpha G(E_{\alpha i}+\mbox{i} \varepsilon )|\Phi_{\alpha i}\rangle .
\end{eqnarray}
Substituting Eq.\ (\ref{g2b}) into (\ref{s12}), 
the first term yields two more terms 
\begin{equation}
S_{\beta j,\alpha i}^{(1)}=\lim\limits_{t\to \infty
}\lim\limits_{\varepsilon \to 0}\mbox{i}\varepsilon 
\mbox{e}^{\mbox{i}
(E_{\beta j}-E_{\alpha i})t}\langle \Phi _{\beta j}|
\widetilde{G}_\alpha
(E_{\alpha i}+\mbox{i}\varepsilon )|\Phi _{\alpha i}\rangle 
\end{equation}
\begin{eqnarray}
S_{\beta j,\alpha i}^{(2)}&=&\lim\limits_{t\to \infty
}\lim\limits_{\varepsilon \to 0}\mbox{i}\varepsilon 
\mbox{e}^{\mbox{i}
(E_{\beta j}-E_{\alpha i})t}\langle \Phi _{\beta j}|
\widetilde{G}_\alpha
(E_{\alpha i}+\mbox{i}\varepsilon ) \nonumber \\
&& U^\alpha G_\alpha ^{(l)}(E_{\alpha i}+\mbox{i}
\varepsilon )|\Phi _{\alpha i}\rangle .
\end{eqnarray}
Using of the properties of the resolvent operators 
the limits can be performed and we arrive at the following, physically 
plausible, result.
The first term, $S_{\beta j,\alpha i}^{(1)}$, is the 
S-matrix of a two-body
single channel scattering on the potential $u_\alpha ^{(l)}$ 
\begin{equation}
S_{\beta j,\alpha i}^{(1)}=\delta _{\beta \alpha }
\delta _{ji}S(u_\alpha ^{(l)}).
\label{s1}
\end{equation}
If $u_\alpha ^{(l)}$ is a pure Coulomb interaction
 $S(u_\alpha ^{(l)})$ falls back
to the S-matrix of the Rutherford scattering, if $u_\alpha ^{(l)}$ is
identically zero
$S_{\beta j,\alpha i}^{(1)}$ equals to unity.
The second term, $S_{\beta j,\alpha i}^{(2)}$, describes a two-body 
multichannel scattering on the
potential $U^\alpha $ 
\begin{equation}
S_{\beta j,\alpha i}^{(2)}=-2\pi \mbox{i}
\delta _{\beta \alpha }\delta
(E_{\beta j}-E_{\alpha i})\langle 
\widetilde{\Phi }_{\beta j}^{(-)}|U^\alpha
|\Phi _{\alpha i}^{(l)(+)}\rangle .  \label{s2}
\end{equation}
The third term gives account of the complete three-body dynamics 
\begin{equation}
S_{\beta j,\alpha i}^{(3)}=-2\pi 
\mbox{i}\delta (E_{\beta j}-E_{\alpha i})\langle \Phi _{\beta j}^{(l)(-)}|V^\alpha |
\Psi _{\alpha i}^{(+)}\rangle .
\label{s3}
\end{equation}

\subsection{Lippmann-Schwinger integral equation for 
$|\Phi _{\alpha}^{(l)}\rangle$}

Starting from the definition of $|\Phi _{\alpha}^{(l)}\rangle$, 
Eq.\ (\ref{phil}), 
by utilizing the resolvent relation (\ref{g2b}) and the definition
(\ref{phitild}), we easily derive a Lippmann-Schwinger equation
\begin{equation}
|\Phi _{\alpha}^{(l)(\pm)}\rangle =| \widetilde{\Phi}_{\alpha}^{(\pm)}\rangle
+ \widetilde{G}_\alpha (E\pm \mbox{i} \epsilon) U^\alpha
|\Phi _{\alpha}^{(l)(\pm)}\rangle,
\label{lsphil}
\end{equation}
where 
$| \widetilde{\Phi}_{\alpha}^{(\pm)}\rangle$ are given by
\begin{equation}
| \widetilde{\Phi}_{\alpha}^{(\pm)}\rangle =
| \widetilde{\chi}_{\alpha}^{(\pm)} \rangle
| \phi_{\alpha} \rangle.  \label{chipm}
\end{equation}
The state $|\widetilde{\chi}_{\alpha}^{(\pm)}\rangle$ is
a scattering state in the Coulomb-like potential $u_\alpha^{(l)}(y_\alpha)$.

\subsection{Faddeev-Merkuriev integral equations 
for the wave function components}

The integral equations for the wave function
$|\Psi _{\alpha}^{(\pm)}\rangle$ are arrived at by combining
the resolvent  relation (\ref{g3b}) and Eq.\ (\ref{Psidef}). In this case
however we have three resolvent relations and therefore we
obtain a triad of Lippmann-Schwinger equations
\begin{eqnarray}
|\Psi _{\alpha}^{(\pm)}\rangle & = |\Phi _{\alpha}^{(l)(\pm)}\rangle +
 & G_\alpha^{(l)} (E \pm {\mathrm{i}} 0)
 V^\alpha |\Psi _{\alpha}^{(\pm)}\rangle \\
|\Psi _{\alpha}^{(\pm)}\rangle  & =  
\phantom{|\Phi _{\alpha}^{(l)}\rangle + \ \ \ \ \  }
& G_\beta^{(l)}(E \pm {\mathrm{i}} 0)
 V^\beta |\Psi _{\alpha}^{(\pm)}\rangle \\
|\Psi _{\alpha}^{(\pm)}\rangle & = 
\phantom{|\Phi _{\alpha}^{(l)}\rangle + \ \ \ \ \  }
& G_\gamma^{(l)}(E \pm {\mathrm{i}} 0)
 V^\gamma |\Psi _{\alpha}^{(\pm)}\rangle .
\label{lstriad}
\end{eqnarray}
Although these three equations together provide unique 
solutions \cite{gloeckle}, their kernels
are not connected therefore they cannot be solved by iterations.
The way out of the problem is to use the Faddeev decomposition 
which leads to equations with connected kernels, thus they are effectively 
Fredholm-type integral equations.

Multiplying each elements of the triad from left by $G^{(l)} v_\alpha^{(s)}$
and utilizing (\ref{fdec}) we get
the set of Faddeev-Merkuriev integral equations for the components
\begin{eqnarray}
|\psi_{\alpha}^{(\pm)} \rangle & = 
|\Phi_{\alpha }^{(l)(\pm)}\rangle + 
& G_\alpha^{(l)} (E \pm {\mathrm{i}} 0)
v^{(s)}_\alpha [ |\psi_{\beta}^{(\pm)} \rangle +
|\psi_{\gamma}^{(\pm)} \rangle  ] \label{fm-eq1} \\ 
|\psi_{\beta}^{(\pm)} \rangle & =  \phantom{
|\Phi_{\beta i}^{(l)(\pm)}\rangle + \ \  }
& G_\beta^{(l)} (E \pm {\mathrm{i}} 0)
v^{(s)}_\beta [ |\psi_{\gamma}^{(\pm)} \rangle +
|\psi_{\alpha}^{(\pm)} \rangle  ] \label{fm-eq2} \\
|\psi_{\gamma}^{(\pm)} \rangle & =  \phantom{
|\Phi_{\beta i}^{(l)(\pm)}\rangle +  \ \   }
& G_\gamma^{(l)} (E \pm {\mathrm{i}} 0)
v^{(s)}_\gamma [ |\psi_{\alpha}^{(\pm)} \rangle +
|\psi_{\beta}^{(\pm)} \rangle  ].
\label{fm-eq3}
\end{eqnarray}
Merkuriev showed that after a certain number of iterations
these equations were reduced to Fredholm integral equations of
the second kind
with compact kernels for all energies, including energies  below
$(E < 0)$ and above $(E  > 0)$ the three-body breakup threshold 
\cite{fm-book}. Thus all the nice properties
of the original Faddeev equations  established for
short-range interactions  remain valid also for the case of Coulomb-like
potentials. We note that 
the triad of Lippmann-Schwinger equations and the set of 
Faddeev equations describe the same physics, the equations have identical
spectra and in fact, the Faddeev equations are the adjoint representations
of the  triad of Lippmann-Schwinger equations \cite{yakovlev}.

Utilizing the properties of the Faddeev components 
the matrix elements in (\ref{s3}) can be rewritten 
in a form  better
suited for numerical calculations 
\begin{equation}
\langle \Phi _{\beta j}^{(l)(-)}|V^\alpha |
\Psi _{\alpha i}^{(+)}\rangle
=\sum_{\gamma \neq \beta }\langle 
\Phi _{\beta j}^{(l)(-)}|v^{(s)}_\beta |\psi_{\gamma i}^{(+)}\rangle .  
\label{s3v}
\end{equation}

Summarizing, in the three-potential formalism, starting from 
$| \widetilde{\Phi}_{\alpha}^{(\pm)}\rangle$, by solving a Lippmann-Schwinger
equation, we determine $|{\Phi}_{\alpha}^{(l)(\pm)}\rangle$. 
Then from $|{\Phi}_{\alpha}^{(l)(+)}\rangle$, by solving the set of
Faddeev-Merkuriev equations, we determine the components 
$|\psi_{\alpha}^{(+)}\rangle$. Finally using Eqs.\ (\ref{s2}) and (\ref{s3v})
we construct the $S$-matrix.

\section{Coulomb-Sturmian separable expansion approach to the
 three-body integral equations }

In order to solve operator equations in quantum mechanics 
one needs a suitable representation for the operators. For solving 
integral equations it is especially advantageous if one uses a 
representation where the Green's operator is simple. 
For the two-body Coulomb Green's operator  there exists a Hilbert-space basis
in which its representation is very simple. This is the Coulomb-Sturmian (CS)
basis. In this representation-space the Coulomb Green's operator can be
given by simple and well-computable analytic functions \cite{papp1}. This
basis forms a countable set. If we represent the interaction term on a finite
subset of the basis it looks like a kind of separable expansion of the
potential, and so the integral equation becomes a set of 
algebraic equations which can then be solved without any further approximation. 
The completeness of the basis ensures the convergence of the method.

This approximation scheme has been thoroughly tested in two-body
calculations. Bound- and resonant-state calculations were presented first
\cite{papp1}. Then the method was extended to 
scattering states \cite{papp2}. Since only 
the asymptotically irrelevant short-range interaction is approximated, 
the correct Coulomb asymptotic is guaranteed \cite{papp3}.
A recent account of this method is presented in Ref.\ \cite{klp}.
The method also proved to be very efficient in solving three-body Faddeev-Noble
integral equations for bound- \cite{pzwp} and scattering-state \cite{pzsc}
problems with repulsive Coulomb interactions.

In subsection A we  define the basis 
states in two- and three-particle Hilbert space. In subsection B 
we review some of the most important formulae of the two-body 
problem. In subsections C and D
we describe the calculation of the $S$-matrix and the
solution of the Faddeev-Merkuriev integral equations.
We follow the line presented in Ref.\ \cite{pzsc}.

\subsection{Basis states}

The Coulomb-Sturmian  functions \cite{rotenberg} in some 
angular momentum state $l$ are defined as 
\begin{equation}
\langle r|nl\rangle =\left[ \frac{n!}{(n+2l+1)!}\right]
^{1/2}(2br)^{l+1} \exp({-br}) L_n^{2l+1}(2br),  \label{basisr}
\end{equation}
$n=0,1,2,\ldots $. Here, $L$ represents the
Laguerre  polynomials and 
$b$ is a fixed parameter.
In an angular momentum subspace they form a complete set 
\begin{equation}
{\bf {1}}=\lim\limits_{N\to \infty }\sum_{n=0}^N|
\widetilde{nl}\rangle
\langle nl|=\lim\limits_{N\to \infty }{\bf {1}}_N,  \label{unity}
\end{equation}
where $|\widetilde{nl}\rangle$ in configuration-space 
representation reads 
$\langle r|\widetilde{nl}\rangle = \langle r|nl\rangle/r$.

The three-body Hilbert space is a direct sum of two-body 
Hilbert spaces.
Thus, the appropriate basis in angular momentum 
representation should be defined as a direct product 
\begin{equation}
| n \nu l \lambda \rangle_\alpha = 
| n l \rangle_\alpha \otimes | \nu
\lambda \rangle_\alpha , \ \ \ \ (n,\nu=0,1,2,\ldots),  
\label{cs3}
\end{equation}
with the CS states of Eq.~(\ref{basisr}). Here $l$
and $\lambda$ denote the angular momenta associated with Jacobi coordinates
$x$ and $y$, respectively. In our three-body Hilbert space basis
we take bipolar harmonics in the
angular variables and CS functions in the radial coordinates.
The completeness relation takes the form (with
angular momentum summation implicitly included) 
\begin{equation}
{\bf 1} =\lim\limits_{N\to\infty} \sum_{n,\nu=0}^N |
 \widetilde{n \nu l
\lambda} \rangle_\alpha \ \mbox{}_\alpha\langle 
{n \nu l \lambda} | =
\lim\limits_{N\to\infty} {\bf 1}_{N}^\alpha,
\end{equation}
where $\langle x_\alpha y_\alpha |
\widetilde{ n \nu l \lambda }
\rangle_\alpha=  
\langle x_\alpha y_\alpha |
{\ n \nu l \lambda }\rangle_\alpha /(x_\alpha y_\alpha)$.
It should be noted that in the three-particle 
Hilbert space we can introduce
three equivalent basis sets which belong to fragmentation 
$\alpha$, $\beta$
and $\gamma$.

\subsection{Coulomb-Sturmian separable expansion in 
two-body scattering problems}

Let us study a two-body case of short-range
 plus Coulomb-like interactions 
\begin{equation}
v_l=v^{(s)}_l+v^C
\end{equation}
and consider the inhomogeneous Lippmann-Schwinger 
equation for the
scattering state $|\psi _l\rangle $ in some partial wave $l$ 
\begin{equation}
|\psi _l\rangle =|\phi _l^C\rangle +g_l^C(E)v^{(s)}_l|\psi _l\rangle .  
\label{LS}
\end{equation}
Here $|\phi _l^C\rangle $ is the regular Coulomb function, 
$g_l^C(E)$ is the
two-body Coulomb Green's operator 
\begin{equation}
g_l^C(E)=(E-h_l^0-v^C)^{-1}
\end{equation}
with the free Hamiltonian $h_l^0$. We make
 the following
approximation on Eq.~(\ref{LS}) 
\begin{equation}
|\psi _l\rangle =|\varphi _l^C\rangle +
g_l^C(E){\bf {1}}_N v^{(s)}_l {\bf {1}}_N |\psi _l\rangle, \label{LSapp}
\end{equation}
i.e.\ we approximate the short-range potential 
$v_l^{(s)} $ by a separable form 
\begin{equation}
v_l^{(s)}=\lim_{N\to\infty} {\bf {1}}_N v_l^{(s)} {\bf {1}}_N \approx
{\bf {1}}_N v_l^{(s)} {\bf {1}}_N 
= \sum_{n, n^{\prime } =0}^N
|\widetilde{
n l}\rangle  \;
\underline{v}_{l}^{(s)}\;\mbox{}
\langle \widetilde{n^{\prime }
l }|  \label{sepfe2b}
\end{equation}
where the matrix
\begin{equation}
\underline{v}_{l_{n n^{\prime }}}^{(s)}=
\langle n l|
v_l^{(s)}| n^{\prime } l \rangle .  
\label{v2b}
\end{equation}
These matrix elements can always be calculated (numerically)
for any reasonable short-range potential. In practice we use
Gauss-Laguerre quadrature, which is well-suited to the CS basis.

Multiplied with the CS states $\langle \widetilde{nl}|$
 from the left, Eq.~(\ref
{LSapp}) turns into a linear system of equations 
for the wave-function
coefficients $\underline{\psi }_{l_n}=\langle 
\widetilde{nl}|\psi _l\rangle $ 
\begin{equation}
\lbrack (\underline{g}_l^C(E))^{-1}-
\underline{v}_l^{(s)}]\underline{\psi }_l=
 (\underline{g}_l^C(E))^{-1}  \underline{\varphi }_l^C,  \label{eq18a}
\end{equation}
where the underlined quantities are matrices with the following elements
\begin{equation}
\underline{\varphi }_{l_n}^C=\langle 
\widetilde{nl}|\varphi _l^C\rangle 
\label{phiov}
\end{equation}
and
\begin{equation}
\underline{g}_{l_{nn^{\prime }}}^C(E)=
\langle \widetilde{nl}|g_l^C(E)|\widetilde{n^{\prime }l}\rangle.  \label{gcme}
\end{equation}

\subsubsection{The matrix elements $\langle \widetilde{nl}|g_l^C(z)|
\widetilde{n^{\prime }l}\rangle$}

The key point in the whole procedure is the exact and analytic
calculation of the CS matrix elements of the Coulomb Green's operator
and of the overlap of the Coulomb and CS functions.
For the Green's matrix we have developed two independent, analytic
approaches. Both are based on the observation that the Coulomb Hamiltonian
possesses an infinite symmetric tridiagonal (Jacobi) matrix structure on
CS basis.

Let us consider the radial Coulomb Hamiltonian 
\begin{equation}
 h^{\rm C}_l=-\frac{\hbar^2}{2m}\left(\frac{\mbox{d}^2 }{\mbox{d} r^2}
- \frac{l(l+1)}{r^2}\right) + \frac{Z}{r}\ ,
\label{coulham}
\end{equation}
where $m$, $l$ and $Z$ stands for the mass, angular momentum and
charge, respectively.
The matrix 
$\underline{J}^{\rm C}_{n n^{\prime}}=
\langle n |(z- h^{\rm C}_l)|n^{\prime} \rangle$ 
possesses a Jacobi structure,
\begin{equation}
\underline{J}^{\rm C}_{nn}=2(n+l+1) (k^2-b^2 )
\frac{\hbar^2}{4mb}- Z 
\label{jii}
\end{equation}
and
\begin{equation}
\underline{J}^{\rm C}_{nn-1}=-[n(n+2l+1)]^{1/2} (k^2+b^2 )
\frac{\hbar^2}{4mb} \ , 
\label{jiip1}
\end{equation}
where $k=(2m z/\hbar^2)^{1/2}$ is the wave number.
The main result of Ref.\ \cite{jmp} is that for Jacobi matrix systems 
the $N$'th leading submatrix 
$\underline{g}^{{\rm C} (N)}_{n n'}$ of the infinite Green's matrix
can be determined by the elements of the Jacobi matrix
\begin{equation}
\label{invn}
\underline{g}^{{\rm C} (N)}_{n n'}=[\underline{J}^{\rm C}_{n n'}+ 
\delta _{n N}\, \, 
\delta _{n' N}\, \, \underline{J}^{\rm C}_{NN+1}\, 
\, C ]^{-1}\ ,
\end{equation}
where $C$ is a  continued fraction
\begin{equation}
C=-\frac{u_N}{d_N+
\frac{\displaystyle u_{N+1}}{\displaystyle d_{N+1}+ 
\frac{\displaystyle u_{N+2}}{ \displaystyle d_{N+2} + \cdots }}} \ ,
\label{frakk}
\end{equation}
with coefficients 
\begin{equation}
u_n=- {\underline{J}^{\rm C}_{n,n-1}}/{\underline{J}^{\rm C}_{n,n+1}}, 
\quad d_n=- {\underline{J}^{\rm C}_{n,n}}/{\underline{J}^{\rm C}_{n,n+1}} 
\ .
\label{egyutth}
\end{equation}

In Ref.\ \cite{jmp} it was shown that although the 
continued fraction $C$ is convergent only on the upper-half $k$ plane 
it can be continued analytically to the whole $k$ plane.
This is because the $u_n$ and $d_n$ coefficients satisfy the limit properties
\begin{eqnarray}
u & \equiv  & \lim_{n\rightarrow \infty }u_n=-1 \\
d & \equiv  & \lim_{n\rightarrow \infty} d_n= 2(k^2 -   b^2)/ 
 ( k^2 +   b^2)\ . 
\label{ud}
\end{eqnarray}  
Then the continued fraction  appears as
\begin{equation}
C=-\frac{u_N}{d_N+
\frac{\displaystyle u_{N+1}}{\displaystyle d_{N+1}+ \cdots +
\frac{\displaystyle u}{ \displaystyle d +  
\frac{\displaystyle u}{ \displaystyle d +    \cdots 
 }}}}\ .
\label{cflim}
\end{equation}
Therefore the tail  $w$ of $C$ satisfies the implicit
relation
\begin{equation}
w=\frac{u}{d+w}\ , 
\label{tail2}
\end{equation}
which is solved by
\begin{equation}
\label{wpm}
w_{\pm}=(b \pm  {\mathrm{i}}k )^2/(b^2+k^2)  \ .
\end{equation}
Replacing the tail of the continued fraction by its explicit analytical 
form $w_{\pm}$, we can speed up the convergence and, more importantly 
turn a non-convergent continued fraction into a
convergent one \cite{lorentzen}. Analytic continuation is achieved
by using $w_{\pm}$  instead of the non-converging tail.
In Ref.\ \cite{jmp} it was 
shown that $w_+$ provides an analytic continuation of the Green's matrix 
to the physical, while $w_{-}$ to the unphysical Riemann-sheet. This way 
Eq.\ (\ref{frakk}) together with (\ref{invn}) provides the CS basis
representation of the Coulomb Green's operator on the whole complex
$k$ plane.  We note here that with the choice 
of $Z=0$ the Coulomb Hamiltonian (\ref{coulham}) reduces to the kinetic 
energy operator and our formulas provide the CS basis representation of the 
Green's operator of the free particle as well.
We emphasize that this procedure 
does not truncate the Coulomb Hamiltonian, because
all the higher $\underline{J}_{nn'}$  matrix elements are implicitly contained 
in the continued fraction. 

We note that $\underline{g}^C$ has already
been calculated before \cite{papp1}. From the J-matrix structure a three-term
recursion relation follows for the matrix elements $\underline{g}^C_{n n'}$.
This recursion relation is solvable if the first element 
$\underline{g}^C_{00}$ is known. 
It is given in a closed analytic form
\begin{eqnarray}
\underline{g}^C_{00}&=& \frac{4mb}{\hbar^2} 
\frac{1}{(b-\mbox{i}k)^2} 
\frac{1}{l+\mbox{i}\eta +1} \nonumber \\
&&\times {_2}F_{1}\left(-l+\mbox{i}\eta,1;l+\mbox{i}\eta+2,
\left(\frac{b+\mbox{i}k} {b-\mbox{i}k}\right)^2 \right),
\label{g00}
\end{eqnarray}
where  $\eta=Z m/(\hbar^2 k)$ is the Coulomb
parameter and ${_2}F_{1}$ is the hypergeometric function. 
For those cases where the first or the second index of
${_2}F_{1}$ is equal to unity, there exists a continued fraction representation,
which is very efficient in practical calculations. It  was shown that 
the two methods lead to numerically identical results for all energies and
our numerical continued fraction representation possesses all the 
analytic properties  of $g^C$. The exact analytic knowledge of  $g^C$ 
allows us to calculate the matrix elements of the full Green's 
operator in the whole complex plane 
\begin{equation}
\underline{g}_l(z)=((\underline{g}_l^C(z))^{-1}-
\underline{v}_l^{(s)})^{-1}.
\label{2bgreen}
\end{equation}

The overlap vector of CS and the Coulomb functions  $\langle 
\widetilde{nl}|\varphi _l^C\rangle$ is known analytically \cite{papp2}.
It can be calculated by a three-term recursion, derived from the
J-matrix, using the starting value
\begin{eqnarray}
\langle 
\widetilde{0l}|\varphi _l^C\rangle &=&
\exp(2\eta\arctan(k/b))\sqrt{\frac{2\pi\eta}{\exp(2\pi\eta)-1 }} \nonumber \\
&&\times \left(\frac{2k/b}{1+k^2/b^2}\right)^{l+1} 
\prod_{i=1}^l \left( \frac{\eta^2+i^2}{i(i+1/2)} \right)^{1/2} .
\end{eqnarray}

\subsection{Calculation of the three-body S-matrix}

The aim of any scattering calculation is to determine the 
S-matrix elements. In our case we need to calculate the terms
(\ref{s1}), (\ref{s2}) and (\ref{s3v}) of the three-potential
picture.

The term $S^{(1)}_{\beta j,\alpha i}$ is trivial because it is just the two-body
S-matrix of the Coulomb-like potential $u_\alpha^{(l)}$.

To calculate the second term, 
$S^{(2)}_{\beta j,\alpha i}$ of Eq.\ (\ref{s2}), 
the matrix elements $\langle \widetilde{\Phi }_{\alpha j}^{(-)}|U^\alpha
|\Phi _{\alpha i}^{(l)(+)}\rangle$ are needed. Since 
$\langle \widetilde{\Phi }_{\alpha j}^{(-)}|$ contains a two-body
bound-state wave function in coordinate $x_\alpha$ this matrix element
is confined to $\Omega_\alpha$, where $U^\alpha$ is of short-range type.
Therefore a separable approximation is justified
\begin{equation}
\langle \widetilde{\Phi }_{\alpha j}^{(-)}|U^\alpha
|\Phi _{\alpha i}^{(l)(+)}\rangle
\approx    \langle \widetilde{\Phi }_{\alpha j}^{(-)}| 
{\bf 1}_N^\alpha U^\alpha {\bf 1}_N^\alpha
|\Phi _{\alpha i}^{(l)(+)}\rangle,
\end{equation}
i.e, in this matrix element, we can approximate $U^\alpha$ by
a separable form
\begin{eqnarray}
U^\alpha   & = &
\lim_{N\to\infty} {\bf 1}_N^\alpha U^\alpha  {\bf 1}_N^\alpha 
\approx {\bf 1}_N^\alpha U^\alpha {\bf 1}_N^\alpha \nonumber \\
  & \approx & \sum_{n,\nu ,n^{\prime },
\nu ^{\prime }=0}^N|\widetilde{n\nu l\lambda }\rangle _\alpha \;
\underline{U}^{\alpha}\;\mbox{}_\alpha \langle \widetilde{n^{\prime }
\nu ^{\prime }l^{\prime }\lambda^{\prime }}|  \label{sepfeU}
\end{eqnarray}
where 
\begin{equation}
\underline{U}^\alpha_{ n\nu l \lambda,
 n^{\prime } \nu^{\prime } l^{\prime} \lambda^{\prime }}=
\mbox{}_\alpha \langle n\nu l\lambda |
U^\alpha |n^{\prime }\nu^{\prime}l^{\prime}{\lambda}^{\prime }
\rangle_\alpha .  
\label{Ua}
\end{equation}
The matrix element appears as
\begin{equation}
\langle \widetilde{\Phi }_{\alpha j}^{(-)}|U^\alpha
|\Phi _{\alpha i}^{(l)(+)}\rangle
\approx \sum^N   \langle \widetilde{\Phi }_{\alpha j}^{(-)}| 
\widetilde{n\nu l\lambda} \rangle_\alpha  \underline{U}^\alpha 
\mbox{}_\alpha\langle \widetilde{n' \nu' l'\lambda'}
|\Phi _{\alpha i}^{(l)(+)}\rangle. \label{sumu}
\end{equation}

In calculating the third term, $S^{(3)}_{\beta j,\alpha i}$ of (\ref{s3v}), 
we have matrix elements of the type 
$\langle \widetilde{\Phi }_{\alpha j}^{l(-)}| v_\alpha^{(s)}  
|\psi_{\beta i}^{(+)}\rangle$. Here 
we can again approximate the short-range potential 
$v_\alpha^{(s)} $ in the three-body
Hilbert space by a separable form 
\begin{eqnarray}
v_\alpha^{(s)}   & = &
\lim_{N\to\infty} {\bf 1}_N^\alpha v_\alpha^{(s)}  {\bf 1}_N^\beta 
\approx {\bf 1}_N^\alpha v_\alpha^{(s)}  {\bf 1}_N^\beta \nonumber \\
  & \approx & \sum_{n,\nu ,n^{\prime },
\nu ^{\prime }=0}^N|\widetilde{
n\nu l\lambda }\rangle _\alpha \;
\underline{v}_{\alpha \beta }^{(s)}\;\mbox{}_\beta \langle 
\widetilde{n^{\prime } \nu ^{\prime }l^{\prime }\lambda
^{\prime }}|  \label{sepfe}
\end{eqnarray}
where 
\begin{equation}
\underline{v}_{{\alpha \beta}_{n \nu l \lambda, 
n^{\prime } \nu ^{\prime } l^{\prime } \lambda^{\prime } }}^{(s)}=
\mbox{}_\alpha \langle n\nu l\lambda |
v_\alpha^{(s)}|n^{\prime }\nu ^{\prime}
l^{\prime }{\lambda }^{\prime }\rangle_\beta .  
\label{vab}
\end{equation}
In (\ref{sepfe}) the ket and bra states belong to different 
fragmentations depending on the
neighbors of the potential operators in the matrix elements.
Finally, the matrix elements take the form
\begin{equation}
\langle {\Phi }_{\alpha j}^{l(-)}| v_\alpha^{(s)}  
|\psi_{\beta i}^{(+)}\rangle \approx
\sum^N 
\langle {\Phi }_{\alpha j}^{l(-)}| 
\widetilde{ n\nu  l \lambda}\rangle_\alpha 
\;\underline{v}_{\alpha \beta }^{(s)}\;
\mbox{}_\beta\langle  \widetilde{ n'\nu'  l' \lambda'} 
 |\psi_{\beta i}^{(+)}\rangle. \label{sumv}
\end{equation}

We conclude that to calculate the S-matrix of the
three-potential formulae we need the CS matrix elements (\ref{Ua})
and (\ref{vab}), which can always be evaluated numerically
by using the transformation of Jacobi coordinates \cite{bb}.
In addition we need the CS wave function components 
$\mbox{}_\alpha\langle \widetilde{n\nu l\lambda} 
| \widetilde{\Phi }_{\alpha i}^{(\pm)}\rangle$, 
$\mbox{}_\alpha\langle \widetilde{n\nu l\lambda} 
| {\Phi}_{\alpha i}^{l(\pm)}\rangle$ and
$\mbox{}_\alpha \langle  \widetilde{ n\nu l \lambda} 
 |\psi_{\alpha }^{(+)}\rangle$. We  determine them in the
following section by solving Lippmann-Schwinger and
Faddeev-Merkuriev integral equations.

It should be noted that  the approximations
(\ref{sepfeU}) and (\ref{sepfe}) used
in calculating the matrix elements (\ref{sumu}) and (\ref{sumv})
become equalities as $N$ goes to infinity.
In practical calculations we increase $N$ until we observe 
numerical convergence in scattering observables.

\subsection{Solution of the three-body integral equations}

In the set of Faddeev-Merkuriev equations (\ref{fm-eq1}-\ref{fm-eq3}) 
we make the approximation of (\ref{sepfe})
\begin{eqnarray}
|\psi _\alpha \rangle & =|\Phi _{\alpha i}^{(l)}\rangle
+ & G_\alpha^{(l)}[{\bf 1}_N^\alpha v^{(s)}_\alpha 
{\bf 1}_N^\beta |\psi _\beta \rangle +
{\bf 1}_N^\alpha v^{(s)}_\alpha {\bf 1}_N^\gamma 
|\psi _\gamma \rangle ]  \label{feqsapp1} \\
|\psi _\beta \rangle & =\phantom{|\Phi _{\alpha i}^{(l)}\rangle
+ \  } & G_\beta^{(l)}[{\bf 1}_N^\beta v^{(s)}_\beta 
{\bf 1}_N^\gamma |\psi _\gamma \rangle +
{\bf 1}_N^\beta v^{(s)}_\beta {\bf 1}_N^\alpha 
|\psi _\alpha \rangle ] \label{feqsapp2} \\
|\psi _\gamma \rangle & =\phantom{|\Phi _{\alpha i}^{(l)}\rangle
+ \  } & G_\gamma^{(l)}[{\bf 1}_N^\gamma v^{(s)}_\gamma 
{\bf 1}_N^\alpha |\psi_\alpha \rangle +
{\bf 1}_N^\gamma v^{(s)}_\gamma {\bf 1}_N^\beta 
|\psi_\beta \rangle ].
\label{feqsapp3}
\end{eqnarray}
Multiplied  by the CS states 
$\mbox{}_\alpha \langle \widetilde{n\nu l\lambda }|$, 
$\mbox{}_\beta \langle \widetilde{n\nu l\lambda }|$ and
$\mbox{}_\gamma \langle \widetilde{n\nu l\lambda }|$, respectively, 
from the left the set of integral equations
turn into a linear system of algebraic equations
for the coefficients of the Faddeev 
components $\underline{\psi }_{\alpha_{ n\nu l
\lambda} }=\mbox{}_\alpha \langle 
\widetilde{n\nu l\lambda }
|\psi _\alpha \rangle $: 
\begin{equation}
\lbrack (\underline{G}^{(l)})^{-1}-
\underline{v}^{(s)}]\underline{\psi }= (\underline{G}^{(l)})^{-1}  
\underline{\Phi }^{(l)},  \label{fep1}
\end{equation}
with 
\begin{equation}
\underline{G}_{\alpha_{ n \nu l \lambda, n^{\prime} \nu^{\prime }
l^{\prime } \lambda^{\prime }}}^{(l)}=
\ \mbox{}_\alpha \langle \widetilde{n\nu l\lambda }|
G_\alpha ^{(l)}|\widetilde{
n^{\prime }\nu ^{\prime }{l^{\prime }}
{\lambda ^{\prime }}}\rangle _\alpha ,
\label{G}
\end{equation}
and 
\begin{equation}
\underline{\Phi }_{\alpha_{ n \nu l \lambda}}^{(l)}=
\mbox{}_\alpha \langle 
\widetilde{n\nu l\lambda }|\Phi _\alpha ^{(l)}\rangle . 
 \label{P}
\end{equation}
Notice that the matrix elements of the Green's 
operator are needed only
between the same partition $\alpha $ whereas 
the matrix elements of the
potentials occur only between different 
partitions $\alpha $ and $\beta $.

\subsubsection{The matrix elements 
$ \mbox{}_\alpha \langle \widetilde{n\nu l\lambda }|
G_\alpha ^{(l)}|\widetilde{n^{\prime }\nu ^{\prime }{l^{\prime }}
{\lambda ^{\prime }}}\rangle _\alpha$ and $\mbox{}_\alpha \langle 
\widetilde{n\nu l\lambda }|\Phi _\alpha ^{(l)}\rangle$}

Unfortunately neither the matrix elements (\ref{G}) 
nor the overlaps (\ref{P}) are known.
The appropriate Lippmann-Schwinger equation for $G_\alpha^{(l)}$ 
was proposed by Merkuriev \cite{fm-book}
\begin{equation}
G_\alpha^{(l)}(z)=G_\alpha^{as}(z) + G_\alpha^{as}(z) V^{as}_\alpha
G_\alpha^{(l)}(z),
\label{LSass}
\end{equation}
where $G_\alpha^{as}$ and $V^{as}_\alpha$ 
are the asymptotic channel Green's operator and potential, respectively. 
A similar equation is valid for $|\Phi _\alpha ^{(l)}\rangle$
\begin{equation}
|\Phi _\alpha ^{(l)}\rangle=|\Phi_\alpha^{as}\rangle
 + G_\alpha^{as}(z) V^{as}_\alpha |\Phi_\alpha^{(l)}\rangle.
\label{LSassphi}
\end{equation}
Both $G_\alpha^{(l)}$ and $|\Phi _{\alpha}^{(l)}\rangle$
are genuine three-body quantities. One may wonder why a
single Lippmann-Schwinger equation suffices. The Hamiltonian $H_\alpha^{(l)}$
has a peculiar property - it has only $\alpha$-type
two-body asymptotic channels. For such systems a single Lippmann-Schwinger
equation provides a unique solution \cite{sandhas}.

The objects $G_\alpha^{as}$, $V^{as}_\alpha$ and $\Phi^{as}_\alpha$
are very complicated. Their leading order terms were
constructed in configurations space in the different asymptotic regions. 
The potential $V^{as}$,  as $|X|\to \infty $,
decays faster than the Coulomb potential
in all directions of the three-body configuration space:
$ V^{as} \sim {\cal O} (|X|^{-1-\epsilon}),\ \epsilon > 0$ \cite{fm-book}.
Therefore we may express the solutions of Eqs.\ (\ref{LSass})
and (\ref{LSassphi}) formally as 
\begin{equation}
(\underline{G}^{(l)}_\alpha )^{-1}= 
(\underline{{G}}^{as}_\alpha )^{-1} -
\underline{V}^{as}_\alpha
\end{equation}
and 
\begin{equation}
[ (\underline{G}^{as}_\alpha )^{-1} -
\underline{V}^{as}_\alpha ]    \underline{\Phi}_\alpha^{(l)} =  
(\underline{G}^{as}_\alpha )^{-1} 
\underline{\Phi}^{as}_\alpha ,
\label{eqphil1}
\end{equation}
respectively, where 
\begin{equation}
\underline{{G}}^{as}_{\alpha_{ n \nu l \lambda, n^{\prime}\nu^{\prime}
l^{\prime} {\lambda}^{\prime} }} =
 \mbox{}_\alpha\langle n
\nu l \lambda | {G}^{as}_\alpha |
 n^{\prime}\nu^{\prime}l^{\prime}{
\lambda}^{\prime}\rangle_\alpha ,  \label{gasme}
\end{equation}
\begin{equation}
\underline{V}^{as}_{\alpha_{ n \nu l \lambda, n^{\prime}\nu^{\prime}
 l^{\prime} {\lambda}^{\prime}}} =
 \mbox{}_\alpha\langle n
\nu l \lambda | V^{as}_\alpha | n^{\prime}\nu^{\prime}
l^{\prime}{\lambda}
^{\prime}\rangle_\alpha \label{vasme}
\end{equation}
and 
\begin{equation}
\underline{\Phi}^{as}_{\alpha_{ n \nu l \lambda}}  = 
\mbox{}_\alpha\langle \widetilde{ n \nu l \lambda }| 
{\Phi}^{as}_\alpha \rangle.
\end{equation}
Here, ${G}^{as}_\alpha$, ${V}^{as}_\alpha$ and $\Phi^{as}_\alpha$ 
appear between finite number of 
of square-integrable CS states, which confine the domain of integration 
to $\Omega_\alpha$. 
In this region, however, $G_\alpha^{as}$ coincides 
with $\widetilde{G}_\alpha$, $V^{as}_\alpha$ with $U^\alpha$ and
$\Phi^{as}_\alpha$ with $\widetilde{\Phi}_\alpha$ \cite{fm-book}.
Finally we have
\begin{equation}
(\underline{G}^{(l)}_\alpha )^{-1}= 
(\underline{\widetilde{G}}_\alpha )^{-1} -
\underline{U}^\alpha ,
\label{gleq}
\end{equation}
where 
\begin{equation}
\underline{\widetilde{G}}_{\alpha_{ n \nu l \lambda, 
 n^{\prime}\nu^{\prime}l^{\prime} {\lambda}^{\prime}}} =
 \mbox{}_\alpha\langle n
\nu l \lambda | \widetilde{G}_\alpha |
 n^{\prime}\nu^{\prime}l^{\prime}{
\lambda}^{\prime}\rangle_\alpha  \label{gtilde}
\end{equation}
and 
\begin{equation}
\underline{U}^\alpha_{ n \nu l \lambda,
 n^{\prime}\nu^{\prime} l^{\prime} {\lambda}^{\prime}} =
 \mbox{}_\alpha\langle n
\nu l \lambda | U^\alpha | n^{\prime}\nu^{\prime}
l^{\prime}{\lambda}
^{\prime}\rangle_\alpha.
\end{equation}
And in a similar way   
\begin{equation}
[ (\underline{\widetilde{G}}_\alpha )^{-1} -
\underline{U}^\alpha ]    \underline{\Phi}_\alpha^{(l)} =  
(\underline{\widetilde{G}}_\alpha )^{-1} 
\underline{\widetilde{\Phi}}_\alpha ,
\label{eqphil}
\end{equation}
where 
\begin{equation}
\underline{\widetilde{\Phi}}_{\alpha_{ n \nu l \lambda}}  = 
\mbox{}_\alpha\langle \widetilde{ n \nu l \lambda }| 
\widetilde{\Phi}_\alpha \rangle.
\end{equation}

We note that from Eq.\ (\ref{gleq}) 
follows that the left side of Eq.\ (\ref{eqphil}) is just the 
inhomogeneous term of Eq.\ (\ref{fep1}). Both  Eqs.\ (\ref{eqphil})
and (\ref{fep1}) are solved with the same inhomogeneous term.

\subsubsection{The matrix elements 
$ \mbox{}_\alpha\langle n
\nu l \lambda | \widetilde{G}_\alpha |
 n^{\prime}\nu^{\prime}l^{\prime}{
\lambda}^{\prime}\rangle_\alpha $ and  
$\mbox{}_\alpha\langle \widetilde{ n \nu l \lambda }| 
\widetilde{\Phi}_\alpha \rangle$}

The three-particle free Hamiltonian
can be written  as a sum of two-particle
free Hamiltonians 
\begin{equation}
H^0=h_{x_\alpha }^0+h_{y_\alpha }^0.
\end{equation}
Then the Hamiltonian $\widetilde{H}_\alpha$ of Eq.\ (\ref{htilde}) appears as 
a sum of two Hamiltonians 
acting on different coordinates 
\begin{equation}
\widetilde{H}_\alpha =h_{x _\alpha }+h_{y_\alpha },
\end{equation}
with $h_{x_\alpha }=
h_{x_\alpha }^0+v_\alpha^C(x _\alpha )$ and $h_{y_\alpha }=
h_{y_\alpha }^0+u_\alpha^{(l)}(y_\alpha )$, which, of course, commute. 
The state $| \widetilde{\Phi}_\alpha \rangle$, which is an eigenstate
of $\widetilde{H}_\alpha$, is a product of a
two-body bound-state wave function in coordinate 
$x_\alpha$ and a two-body scattering-state wave function in coordinate 
$y_\alpha$. Their CS representations are known from the 
two-particle case described before. 

The matrix elements of $\widetilde{G}_\alpha$ can be determined by 
making use of the convolution theorem 
\begin{eqnarray}
\widetilde{G}_\alpha (z) & = & (z-h_{x_\alpha }-
h_{y_\alpha })^{-1} \nonumber \\
&=& \frac 1{2\pi \mbox{i}}\oint_C dz' (z-z'-h_{x_\alpha })^{-1}  
 (z'-h_{y_\alpha })^{-1}. 
 \label{contourint}
\end{eqnarray}
The contour $C$ should encircle, in positive direction, the 
spectrum of $h_{y_\alpha }$
without penetrating into the spectrum of $h_{x_\alpha }$. 

The convolution theorem follows from a more general formula.
A function of a self adjoint operator $h$ is defined as
\begin{eqnarray}
f(h)=\frac 1{2\pi \mbox{i}} \oint_C dz f(z) (z-h)^{-1},
\end{eqnarray}
where $C$ is a contour around the spectrum of $h$ and $f$ should be
analytic on the region encircled by $C$.

In the following we suppose that $u^{(l)}$ either vanishes or is
a repulsive Coulomb-like potential. This assumption is not necessary but
it greatly simplifies the analysis below. Numerical examples show that
there are a great many physical three-body systems where this condition 
is satisfied. This condition ensures that $h_y$ does not have bound
states.

\begin{figure}
\psfig{file=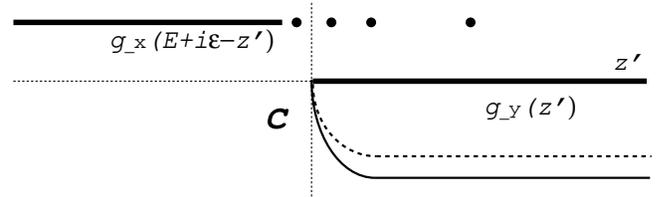,width=8.5cm}

\caption{Analytic structure of $g_{x_\alpha }(z-z^\prime)\;
g_{y_\alpha}(z^\prime)$ as a function of $z^\prime$ with
$z=E+{\mathrm{i}}\varepsilon$, $E<0$, $\varepsilon>0$.
The contour $C$ encircles the continuous spectrum of
$h_{y_\alpha}$. A part of it, which goes on the unphysical
Riemann-sheet of $g_{y_\alpha}$, is drawn by broken line.}

\label{fig1}
\end{figure}

To examine the analytic structure of the integrand (\ref{contourint}) let us
shift the spectrum of $g_{x_\alpha }$ by
taking  $z=E +{\mathrm{i}}\varepsilon$  with
positive $\varepsilon$. In doing so,
the two spectra become well separated and
the spectrum of $g_{y_\alpha}$ can be encircled.
The contour $C$ is deformed analytically
in such a way that the upper part descends to the unphysical
Riemann sheet of $g_{y_\alpha}$, while
the lower part of $C$ can be detoured away from the cut
 [see  Fig.~\ref{fig1}]. The contour still
encircles the branch cut singularity of $g_{y_\alpha}$,
but in the  $\varepsilon\to 0$ limit avoids the singularities of $g_{x_\alpha}$.
 Thus, the mathematical conditions for
the contour integral representation of $\widetilde{G}_\alpha (z)$ in
Eq.~(\ref{contourint}) is met. 
The matrix elements $\underline{\widetilde{G}}_\alpha$
can be cast in the form
\begin{equation}
\widetilde{\underline{G}}_\alpha (z)=
 \frac 1{2\pi \mathrm{i}}\oint_C
dz^\prime \,\underline{g}_{x_\alpha }(z-z^\prime)\;
\underline{g}_{y_\alpha}(z^\prime),
\label{contourint2}
\end{equation}
where the corresponding CS matrix elements of the two-body Green's operators in
the integrand are known analytically for all complex energies.

\section{Test of the method}

We demonstrate the power of this new method by calculating elastic phase shifts of
$e^++H$ scattering below the $Ps(n=1)$ threshold and cross sections
of the $e^++H$ elastic scattering
as well as $p^++Ps$ reaction channels up to the $Ps(n=2)$ threshold. 
In all examples we have total angular momentum $L=0$ and we have taken 
angular momentum channels up to $l=10$. We use atomic units.

Let us numerate the particles $e^+$, $p$ and $e^-$,
with masses $m_{e^\pm}=1m_e$ and $m_{p}=1836.1527m_e$, by $1$, $2$ and $3$,
respectively. In the channel $3$ there are no two-body
asymptotic channels since the particles $e^+$ and $p$ do not form bound states.
Therefore, we can take $v_3^{(s)}\equiv 0$ and include the total
$v_3^C$ in the long range Hamiltonian
\begin{eqnarray}
H &=& H^{(l)}+v_1^{(s)}+v_2^{(s)}, \\
H^{(l)} &=& H^0+v_1^{(l)}+v_2^{(l)}+v_3^C .
\end{eqnarray}
In this case $|\psi_3 \rangle \equiv 0$ and we have the set of
two-component Faddeev-Merkuriev equations
\begin{eqnarray}
|\psi_1 \rangle &=& |\phi^{(l)}_1\rangle +G^{(l)}_1 v^{(s)}_1 |\psi_1\rangle \\
|\psi_2 \rangle &=& \phantom{|\phi^{(l)}_1\rangle + \  } 
G^{(l)}_2 v^{(s)}_2 |\psi_2\rangle .
\end{eqnarray}
The parameters of the splitting function $\zeta$ of Eq.\ (\ref{oma1})
are rather arbitrary. The final converged results should be insensitive 
to their values; our numerical experiences confirm this expectation. 
For the parameters of $\zeta$ we 
have taken $\nu=2.1$, $x^0=3$ and $y^0=10$, whereas
for the parameters
of CS functions we have taken $b=0.9$. We have experienced that the 
rate of convergence is rather
insensitive on the choice of $b$ over a broad interval.

First we examine the convergence of
the results for cross sections at incident wave numbers
$k_1=0.71$, $k_1=0.75$ and  $k_1=0.8$,
which correspond to  scattering states  in the Ore gap.
Table \ref{tabconv1} shows the convergence of
$e^++H->e^++H$ elastic scattering
($\sigma_{11}$) and $e^++H->p^++Ps$ positronium formation ($\sigma_{12}$)
cross sections (in $\pi a_0^2$) with respect to $N$, the number of CS functions in the
expansion, and with respect to increasing the angular momentum
channels in the bipolar expansion. For comparison we provide the results
of Ref.\ \cite{kwh}. We can see that very good accuracy is 
achieved even with relatively low $N$ in the expansion.

In Table \ref{tabshift} we compare 
our converged results for phase shifts (in radians) 
below the $Ps(n=1)$ threshold to that of other methods.  
Ref.\ \cite{5} is the best variational calculation. In 
Ref.\ \cite{20} the Schr\"odinger equation was solved by means of
finite-element method. In Refs.\ \cite{14} and \cite{kwh} the 
configurations space Faddeev-Merkuriev differential
equations were solved using the bipolar harmonic expansion method 
and in total angular momentum representation, respectively.
We can report perfect agreements with previous calculations.

In Table \ref{4channel} we present partial cross sections 
in the $H(n=2)-Ps(n=2)$ gap (threshold energies 0.7496-0.8745 Ry). 
In Ref.\ \cite{hu99} the configurations space Faddeev-Merkuriev differential
equations were solved using the bipolar harmonic expansion in the angular
variables an quintic spline expansion in the radial coordinates. 
We can report fairly good agreements.

\section{Conclusion}

We have extended the three-potential formalism for 
treating the three-body scattering problem with all kinds of
Coulomb interactions including attractive ones.
We adopted Merkuriev's approach and split the Coulomb potentials
in the three-body configuration space into short-range and long-range
terms. In this picture the three-body Coulomb
scattering process can be decomposed into a 
single channel Coulomb scattering, a two-body
multichannel scattering on the intermediate-range 
polarization potential and a genuinely three-body scattering due to the
short-range potentials. The formalism provides us a set of 
Lippmann--Schwinger  and Faddeev-Merkuriev integral equations.

These integral equations are certainly too complicated for the most of the
numerical methods available in the literature. The Coulomb-Sturmian
separable expansion method can be successfully applied.
 It solves the three-body
integral equations by expanding only the short-range terms
in a separable form on Coulomb-Sturmian basis 
while treating the long-range terms in an exact manner via a proper integral 
representation of the three-body channel distorted Coulomb Green's operator.
The use of the Coulomb-Sturmian basis 
is essential as it allows an exact analytic representation of the two-body 
Green's operator, and thus the contour integral for the channel distorted 
Coulomb Green's operator can be calculated.
The method provides solutions which are
asymptotically correct, at least in $\Omega_\alpha$, which
is sufficient if the scattering process starts from a two-body
asymptotic state. Since the two-body Coulomb Green's operator 
is exactly calculated all  thresholds are automatically in the right location
irrespective of the rank of the separable approximation.
The method possesses good convergence properties and in 
practice it can be made arbitrarily accurate by employing an increasing 
number of terms in the expansion. Certainly, there is plenty of room
for improvement but we are convinced that this method can be a very 
powerful tool for studying three-body systems with Coulomb interactions.

\acknowledgements
This work has been supported by the NSF Grant No.Phy-0088936
and by the OTKA Grant No.\ T026233. We also acknowledge the
generous allocation of computer time at the NPACI, formerly 
San Diego Supercomputing Center, by the National Resource Allocation
Committee and at the Department of Aerospace Engineering
of CSULB.

\newpage
\begin{table}[tbp]
\caption{Convergence of
$e^++H->e^++H$ elastic scattering
($\sigma_{11}$) and $e^++H->p+Ps$ positronium formation ($\sigma_{12}$)
cross sections (in $\pi a_0^2$) with respect to $N$, the number of CS functions in the
expansion, and with respect to increasing the angular momentum
channels ($l_{max}$) in the bipolar basis.}
\label{tabconv1}
\begin{tabular}{|l|cc|cc|cc|}
   & \multicolumn{2}{c|}{$l_{max}=6$} & \multicolumn{2}{c|}{$l_{max}=8$} 
   & \multicolumn{2}{c|}{$l_{max}=10$}  \\
$N$& $\sigma_{11}$& $\sigma_{12}$& $\sigma_{11}$& $\sigma_{12}$& 
$\sigma_{11}$& $\sigma_{12}$  \\ \hline
\multicolumn{7}{|c|}{$k_1=0.71$, 
Ref.\ \cite{kwh}: $\sigma_{11}=0.025$, $\sigma_{12}=0.0038$} \\ \hline
12 & 0.02662 & 0.00423 & 0.02664 & 0.00397 & 0.02665 & 0.00393 \\
13 & 0.02608 & 0.00424 & 0.02609 & 0.00398 & 0.02610 & 0.00394 \\
14 & 0.02581 & 0.00423 & 0.02582 & 0.00398 & 0.02583 & 0.00394 \\
15 & 0.02562 & 0.00424 & 0.02561 & 0.00398 & 0.02562 & 0.00395 \\
16 & 0.02548 & 0.00425 & 0.02546 & 0.00400 & 0.02547 & 0.00396 \\
17 & 0.02541 & 0.00426 & 0.02539 & 0.00401 & 0.02539 & 0.00397 \\
18 & 0.02532 & 0.00427 & 0.02529 & 0.00401 & 0.02530 & 0.00398 \\
19 & 0.02528 & 0.00427 & 0.02524 & 0.00402 & 0.02525 & 0.00398 \\
20 & 0.02522 & 0.00428 & 0.02517 & 0.00403 & 0.02518 & 0.00399 \\
\hline
\multicolumn{7}{|c|}{$k_1=0.75$, 
Ref.\ \cite{kwh}: $\sigma_{11}=0.044$, $\sigma_{12}=0.0043$} \\ \hline
12 & 0.04412 & 0.00441 & 0.04412 & 0.00424 & 0.04413 & 0.00422 \\
13 & 0.04345 & 0.00440 & 0.04344 & 0.00422 & 0.04345 & 0.00421 \\
14 & 0.04318 & 0.00440 & 0.04317 & 0.00423 & 0.04318 & 0.00421 \\
15 & 0.04280 & 0.00440 & 0.04278 & 0.00423 & 0.04279 & 0.00421 \\
16 & 0.04269 & 0.00440 & 0.04265 & 0.00423 & 0.04266 & 0.00422 \\
17 & 0.04252 & 0.00441 & 0.04248 & 0.00424 & 0.04249 & 0.00423 \\
18 & 0.04246 & 0.00442 & 0.04240 & 0.00425 & 0.04241 & 0.00423 \\
19 & 0.04238 & 0.00442 & 0.04232 & 0.00426 & 0.04232 & 0.00424 \\
20 & 0.04232 & 0.00442 & 0.04225 & 0.00426 & 0.04226 & 0.00424 \\
\hline
\multicolumn{7}{|c|}{$k_1=0.80$, 
Ref.\ \cite{kwh}: $\sigma_{11}=0.063$, $\sigma_{12}=0.0047$} \\ \hline
12 & 0.06572 & 0.00475 & 0.06571 & 0.00467 & 0.06572 & 0.00467 \\
13 & 0.06573 & 0.00481 & 0.06571 & 0.00473 & 0.06572 & 0.00473 \\
14 & 0.06518 & 0.00483 & 0.06515 & 0.00475 & 0.06517 & 0.00475 \\
15 & 0.06488 & 0.00485 & 0.06484 & 0.00477 & 0.06486 & 0.00477 \\
16 & 0.06457 & 0.00486 & 0.06452 & 0.00478 & 0.06453 & 0.00478 \\
17 & 0.06440 & 0.00487 & 0.06433 & 0.00479 & 0.06435 & 0.00479 \\
18 & 0.06427 & 0.00487 & 0.06420 & 0.00479 & 0.06422 & 0.00480 \\
19 & 0.06418 & 0.00487 & 0.06409 & 0.00480 & 0.06411 & 0.00480 \\
20 & 0.06412 & 0.00488 & 0.06402 & 0.00480 & 0.06404 & 0.00480
\end{tabular}
\end{table}

\begin{table}[tbp]
\caption{Phase shifts (in radians) 
of $e^++H->e^++H$ elastic scattering below
the positronium formation threshold.}
\label{tabshift}
\begin{tabular}{lccccc}
$k$&  Ref.\ \cite{5} &  Ref.\ \cite{20} &  Ref.\ \cite{14} & 
Ref.\ \cite{kwh} & This work \\ \hline
0.1 &  0.1483 & 0.152 & 0.149 &  0.149 &  0.1480 \\
0.2 &  0.1877 & 0.188 & 0.188 &  0.189 &  0.1876 \\
0.3 &  0.1677 & 0.166 & 0.166 &  0.169 &  0.1673 \\
0.4 &  0.1201 & 0.118 & 0.120 &  0.121 &  0.1199 \\
0.5 &  0.0624 & 0.061 & 0.060 &  0.062 &  0.0625 \\
0.6 &  0.0039 & 0.003 &       &  0.003 &  0.0038 \\
0.7 & -0.0512 &-0.053 &       & -0.050 & -0.0513
\end{tabular}
\end{table}

\begin{table}[tbp]
\caption{Partial cross sections (in $\pi a_0^2$) 
in the $H(n=2)-Ps(n=2)$ gap
(threshold energies 0.7496-8745 Ry). Numbers $1$,$2$,$3$ and $4$ denote the
channels $e^++H(1s)$, $e^++H(2s)$, $e^++H(2p)$ and  $p^++Ps(1s)$,
respectively.}
\label{4channel}
\begin{tabular}{lccccc}
$E_1$(Ry) &  & $\sigma_{11}$  & $\sigma_{12}$  & $\sigma_{13}$   
 & $\sigma_{14}$\\ \hline
0.77 & Ref.\ \cite{hu99} & 0.090  & 0.000702 & 0.000454 & 0.00572 \\
0.77 & This work         & 0.0951 & 0.000673 & 0.000331 & 0.00558 \\ \hline
0.80 & Ref.\ \cite{hu99} & 0.096  & 0.00115  & 0.000364 & 0.00585 \\
0.80 & This work         & 0.1010 & 0.00127  & 0.000371 & 0.00563 \\ \hline
0.83 & Ref.\ \cite{hu99} & 0.0993 & 0.00170  & 0.000885 & 0.00581 \\
0.83 & This work         & 0.1063 & 0.00163  & 0.000813 & 0.00566 \\ \hline
0.84 & Ref.\ \cite{hu99} & 0.101  & 0.00190  & 0.00113 & 0.00580 \\
0.84 & This work         & 0.1080 & 0.00173  & 0.00105 & 0.00566 
\end{tabular}
\end{table}

\end{document}